# SAM2-Aug: Prior knowledge-based Augmentation for Target Volume Auto-Segmentation in Adaptive Radiation Therapy Using Segment Anything Model 2

Guoping Xu, Yan Dai, Hengrui Zhao, Ying Zhang, Jie Deng, Weiguo Lu, You Zhang

The Medical Artificial Intelligence and Automation (MAIA) Laboratory

Department of Radiation Oncology

University of Texas Southwestern Medical Center, Dallas, TX 75390, USA

Corresponding address:

You Zhang

Department of Radiation Oncology

University of Texas Southwestern Medical Center

2280 Inwood Road

Dallas, TX 75390

Email: You.Zhang@UTSouthwestern.edu

Tel: (214) 645-2699

## Abstract

**Purpose:** Accurate tumor segmentation is essential for adaptive radiation therapy (ART), yet remains a time-consuming and labor-intensive process with considerable inter-user variations. Recent advancements in foundation models, such as the Segment Anything Model 2 (SAM2), offer strong potential for prompt-based segmentation while still facing accuracy challenges in tumor segmentation. This study proposes novel prior knowledge-based augmentation strategies to improve SAM2's performance for tumor segmentation in the ART workflow.

**Methods and Materials:** We propose two prior knowledge-based augmentation strategies for improving the tumor segmentation performance of SAM2 in MR-LINAC-based ART: (1) leveraging prior MR images and their corresponding annotations to provide contextual guidance, and (2) enhancing prompt robustness through random bounding box expansion and mask erosion/dilation during model fine-tuning. The SAM2-Aug model was fine-tuned and tested using a One-Seq-Liver dataset (115 MRI scans of the same imaging sequence from 31 liver cancer patients, simulation scans not counted). In addition, the fine-tuned model was directly evaluated on two other datasets: Mix-Seq-Abdomen (88 MRI scans of mixed sequences from 28 patients with abdominal tumors, simulation scans not counted) and Mix-Seq-Brain (86 MRI scans of mixed sequences from 37 patients with brain tumors, simulation scans not counted).

**Results:** SAM2-Aug outperformed state-of-the-art convolutional, transformer-based, and prompt-driven segmentation models across all datasets. It achieved mean(±s.d.) Dice scores of 0.86±0.08, 0.89±0.06, and 0.90±0.07 for ITV (liver), ITV (abdomen), and CTV (brain), respectively. SAM2-Aug demonstrated robustness to variable tumor boundaries, and generalized across modalities—without the need for large-scale retraining. The integration of prior context and prompt augmentation significantly improved segmentation quality in challenging ART scenarios.

**Conclusions:** By incorporating prior imaging information and enhancing prompt diversity, SAM2-Aug delivers accurate, robust, and generalizable tumor segmentation. These advancements offer a potential path toward more efficient and precise ART workflows. The code and trained models will be made publicly available at https://github.com/apple1986/SAM2-Aug.

## 1. Introduction

Radiation therapy aims to deliver a prescribed radiation dose to tumor targets while minimizing exposure to surrounding normal tissues [1, 2]. It is a highly effective treatment; however, its success relies on accurate tumor localization and delineation. Tumor delineation, mostly manual, is time-consuming, labor-intensive, and prone to inter-user variability. Though substantial advancements have been achieved in normal tissue/organ automatic contouring, automatic segmentation for tumors remains a substantial challenge [3, 4]. Adaptive radiation therapy (ART), which allows the re-optimization of treatment plans based on anatomical changes observed in daily treatment imaging, enables more precise treatments but requires the re-delineation of tumors, adding a substantial burden on clinical workflows. One potential advantage of ART, however, is the availability of prior information (images/segmentations from treatment simulation or prior treatment adaptations). Leveraging such information can potentially reduce manual annotation efforts and automate the ART workflow [5].

In recent years, deep learning (DL) methods [6, 7] trained on large datasets have demonstrated state-of-the-art performance in many segmentation tasks [8-11]. Among these methods, the U-Net architecture has garnered substantial attention and success due to its simplicity and effectiveness [12, 13]. U-Net consists of an encoder, a decoder, and skip connections that bridge the encoder and decoder. The encoder extracts semantic and contextual features, while the decoder progressively upsamples feature maps to produce finer segmentation results. Skip connections transfer localization information from the encoder to the decoder, facilitating accurate object segmentation. Numerous U-Net-based variants have been proposed for medical image segmentation, including UNet++ [14], Swin UNETR [15], UNETR++ [16], MedNeXt [17] and nnU-Net [18]. These methods explore multi-scale feature fusion using skip connections or pyramid pooling (as in UNet++ [14], DeepLab [9] and SegFormer [19]), local and global feature learning based on convolution and transformer operations (as in SegResNet [20] and TransUNet [11]), or efficient segmentation through novel convolutional operations or hybrid convolution-transformer architectures (as in ConvUNeXt [21] and LeViT-UNet [10]). Despite significant advancements in task-specific DL methods for medical image segmentation, these approaches often lack generalizability across diverse medical imaging modalities [22]. Additionally, most existing methods primarily focus on organ segmentation, such as liver [23], heart [11], kidneys [24], and lungs [25]. In contrast, tumor segmentation remains challenging due to the substantial variability in tumor shapes/intensities/textures, the low contrast between tumors and surrounding tissues or organs, as well as the heterogeneous nature of medical images [26]. As shown in Figure 1, the liver tumors (highlighted with red contours) can be difficult to differentiate from the background even for MRI scans. These limitations hinder the clinical application of such methods for radiotherapy, where accurate tumor delineation is essential to ensure the effective treatment [27].

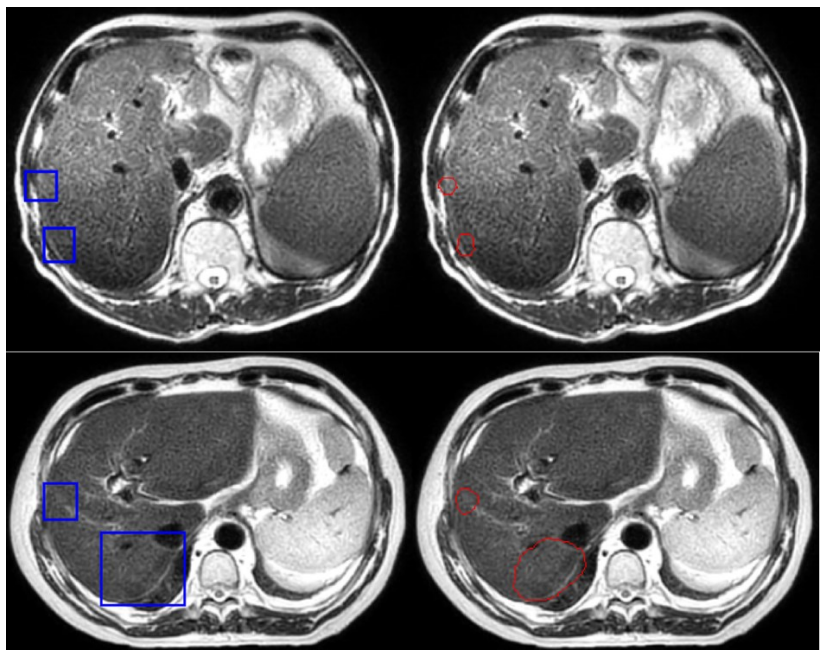

**Figure 1**. Liver tumors highlighted with red contours on MR images for two cases (right). For better visualization against the background, a blue bounding box expanded by 10 pixels in all four directions is also shown for each case (left). As demonstrated, accurate tumor delineation remains challenging.

Unlike previous task-specific and modality-specific DL-based approaches, a foundation model known as Segment Anything Model (SAM) was introduced for segmentation tasks [28]. SAM employs a novel prompt-based segmentation framework for generic object segmentation. Trained on 11 million 2D natural images and over 1 billion segmentation masks, SAM demonstrated remarkable segmentation performance when given appropriate prompts, such as points, boxes, masks, or texts, setting a new milestone in segmentation capabilities. Due to its outstanding performance, SAM has inspired numerous adaptations for medical image segmentation, including MedSAM [22], LeSAM [26], SAM-Med2D [29], and SAM-Med3D[30]. These SAM-based methods were fine-tuned on large-scale medical datasets and demonstrated better accuracy and generalizability compared to modality-specific specialist models. Recently, SAM 2 was introduced as an advancement over its predecessor, designed for both image and video segmentation. It delivers enhanced accuracy and achieves six times faster performance for image segmentation tasks by utilizing streaming memory for real-time video processing [29]. Building on the capabilities of SAM 2, MedSAM-2 was developed to treat 3D medical images as video data (rolling along the slice direction), demonstrating state-of-the-art performance compared to both task-specific and prompt-based models across a wide range of organ and lesion segmentation datasets [31].

Although SAM2 and its variants [32-34] demonstrate robust 2D/3D medical image segmentation performance, they usually require re-training on large-scale medical image datasets with corresponding labels, which poses a significant challenge in scenarios like ART, where the available datasets are typically small. In addition, the unique workflow of ART also brings some additional opportunities for potential segmentation enhancement: for each adaptive fraction, there is almost always some prior information available. Such information includes the images/segmentation from the original treatment simulation or from prior adaptive fractions, which may provide valuable contextual knowledge for current tumor segmentation. To enhance the applicability of SAM2-based methods for tumor segmentation in ART, in this study we proposed an innovative approach that focuses on integrating prior information, enhancing prompt augmentation, and fine-tuning SAM2 for ART. The main contributions of this study are as follows:

(1) We introduced prior MR images and their annotations as input to SAM2, leveraging previous information for augmentation to improve current tumor segmentation accuracy.

(2) We propose a novel prompt augmentation strategy that involves randomly dropping out bounding boxes or masks, as well as dynamically expanding bounding boxes or eroding/dilating masks during training. This approach aims to enhance the model's robustness and effectiveness for tumor segmentation.

(3) We investigate various fine-tuning strategies for SAM2 and evaluate the proposed methods on three distinct tumor segmentation datasets, demonstrating superior performance compared to existing task-specific convolutional/transformer-based models and prompt-specific SAM-based interactive approaches.

## 2. Related works

*2.1 Data Augmentation*

Deep learning often requires large, annotated training datasets, which are limited due to cost, scarcity, and privacy concerns [35]. Data augmentation helps alleviate this by synthetically increasing data diversity [36]. Traditional techniques—e.g., rotation, flipping, scaling, and noise addition—improve generalizability but may fail to capture the anatomical complexity and contextual information essential for tasks like tumor segmentation. Advanced data augmentation strategies are increasingly being adopted, typically falling into two categories: (1)

multimodal image integration and (2) region-specific feature or prior knowledge enhancement. For instance, DiSegNet [37] utilizes both PET and CT imaging to segment pathological lymph nodes, leveraging PET for localization and CT for structural detail. Similarly, multimodal imaging combination strategies involving CT, PET, and MRI have demonstrated superior performance over single-modality methods in segmenting head and neck squamous cell carcinoma [38].

In contrast to direct image integration, several approaches in ART focus on feature-level augmentation targeting specific regions or locations. For example, Ren et al. [39] employed gradient maps of tumor lesions and prior segmentations from earlier radiotherapy sessions to guide current tumor boundary delineation in head and neck cancer. Similarly, patient-specific clinical knowledge—derived from previous clinical target volume (CTV) contours and population-based mesorectal deformation patterns—was incorporated in rectal cancer segmentation, demonstrating improved performance over MRI-only models [40]. In another study, a tailored LSTM-UNet architecture was trained using both prior cone-beam computed tomography (CBCT) images with contours and current CBCT scans as dual-branch input, effectively leveraging historical fraction information for enhanced segmentation [41].

Inspired by the two categories of data augmentation strategies, we integrate prior MR images and their corresponding annotations with the current MR scan as input. This design provides both temporal and anatomical context, leveraging prior delineation knowledge to improve tumor segmentation in ART. Despite its simplicity, this approach is highly effective when combined with the SAM2 framework, owing to SAM2's robust feature representation capabilities.

*2.2 Prompt augmentation for SAM-based methods*

Prompt augmentation, unlike traditional data augmentation, enhances the structure and diversity of input prompts rather than image contents or features. By incorporating user-provided prompts (like points, bounding boxes, and masks), integrative image segmentation can obtain more precise and targeted results for SAM-based frameworks. However, simple prompts may lead to ambiguous results [28]. In SAMAug [42], four different point augmentation strategies, including random sampling, sampling based on maximum difference entropy, maximum distance, and saliency, were proposed to improve image segmentation performance. In SAMatch, varying points or bounding boxes are automatically generated from a U-Net-based model under training for input into SAM [43]. In addition, prompt-learning-based augmentation methods have been explored to enhance the performance of SAM-based methods in various domains [44]. For instance, a spatial-semantic prompt learning approach was proposed in [45], which jointly optimizes spatial and semantic prompts in the embedding space and selectively leverages knowledge from pre-trained prompt encoders, demonstrating effectiveness across diverse datasets. Similarly, an adaptive prompt learning module was introduced in [46], which generates scale- and shape-adaptive visual prompts and incorporates a multi-source, multi-level mask decoder to improve segmentation precision based on SAM.

In ART, prior treatment annotations offer critical insights into tumor size, shape, and location. As shown in Figure 2, there is a strong spatial and morphological correlation between previous and current tumors, underscoring the potential of leveraging historical annotations to guide more accurate segmentation. To bridge this gap, we introduce a novel prompt augmentation strategy that integrates current bounding box prompts with prior mask prompts derived from earlier annotations. This combination is further refined through stochastic dropout and morphological operations (e.g., dilation and erosion), enhancing robustness to imperfect prompts and improving SAM2's adaptability to the precision requirements of ART.

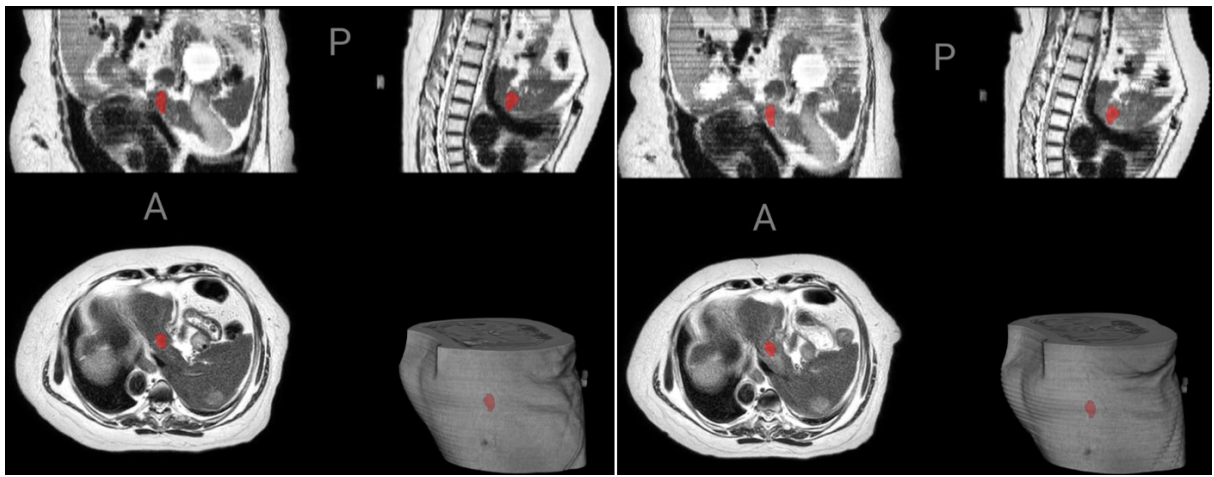

**Figure 2.** Illustration of a patient's prior MRI with tumor annotation (left panel), alongside the current MRI with tumor annotation (right panel). It highlights the strong correlation between the previous and current tumors in terms of size, shape, and location, demonstrating the potential value of leveraging prior information for improved tumor segmentation.

*2.3 Finetuning SAM-based methods for medical image segmentation*

Fine-tuning SAM-based methods has proven an effective strategy for transferring knowledge from natural image domains to medical image segmentation tasks. In MedSAM [22], all trainable parameters of SAM's image encoder and mask decoder were fine-tuned using a curated large-scale medical image dataset, comprising over 1.5 million image-mask pairs spanning 10 imaging modalities. Alternatively, MedSAM-Adapter [34], SAM2-Adapter [33] and DD-Adapter [47] introduce adapter modules integrated into the SAM or SAM2 architecture, enabling task-specific adaptation while retaining the benefits of pre-trained weights. MA-SAM [48] injects a series of 3D adapters into the image encoder of SAM and updates only a small portion of weights with a parameter-efficient fine-tuning strategy [49], aiming to incorporate third-dimensional volumetric information for medical image segmentation.

In contrast to these approaches, which require either large-scale datasets or architectural modifications, we demonstrate that directly fine-tuning the image encoder, prompt encoder, and mask decoder of SAM2 using a relatively small dataset—when combined with our proposed data and prompt augmentation strategies—offers a more effective and streamlined solution for tumor segmentation in ART.

**3. Methods**

In this work, we propose SAM2-Aug, a comprehensive framework that enhances SAM2 through generalized data and prompt augmentations, coupled with parameter fine-tuning for tumor segmentation in ART. The overall architecture of SAM2-Aug is illustrated in Figure 3.

Built upon the SAM2 backbone, SAM2-Aug introduces three key components to enhance segmentation accuracy in ART: (1) *Data Augmentation* – integrating prior MR images and tumor annotations with the current scan to provide temporal and anatomical context; (2) *Prompt Augmentation* – combining bounding boxes and prior masks with random morphological and dropout operations to improve robustness; and (3) *Fine-Tuning* – adapting SAM2 to ART-specific tasks using a limited set of annotated medical images to address data scarcity in clinical applications. In the following subsections, we detail the preliminaries of SAM2, followed by our proposed data and prompt augmentation strategies and the fine-tuning process.

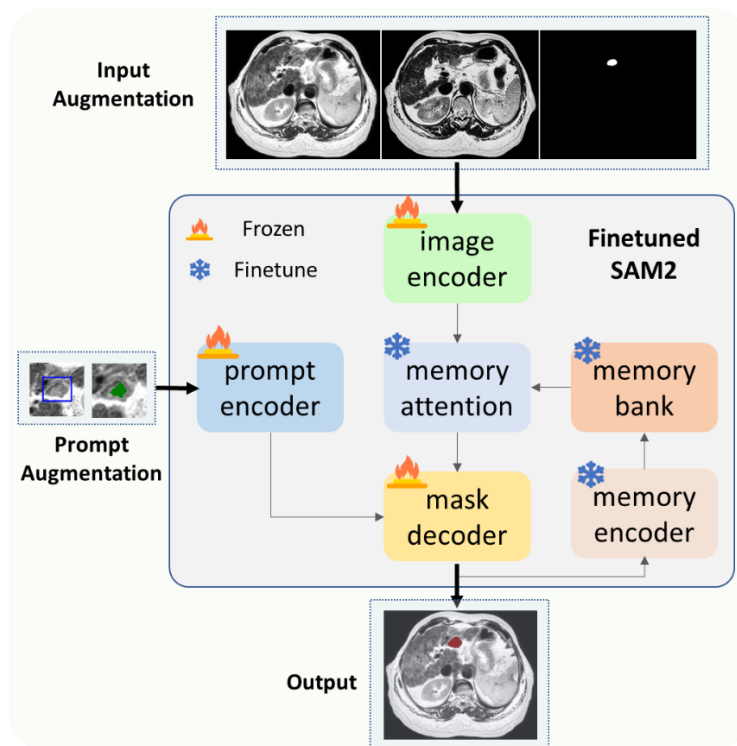

**Figure 3**. The SAM2-Aug framework. Built upon the SAM2 architecture, SAM2-Aug is designed to enhance performance in tumor segmentation for adaptive radiation therapy (ART). The framework incorporates three key components: (1) *Data Augmentation*, achieved by integrating previous MR images and tumor annotations with the current MR image; (2) *Prompt Augmentation*, involving the fusion of bounding boxes and prior masks with dropout and morphological operations to improve robustness; and (3) *Fine-tuning* of the original SAM2 framework using a limited set of annotated data from ART applications.

*3.1 Preliminaries of Segment Anything Models*

Segment Anything Models (SAMs) [28, 50] are foundational models that introduce a novel prompt-based paradigm, where the model generates segmentation masks in response to user-provided prompts. This approach enables interactive segmentation and facilitates large-scale data annotation via model-in-the-loop pipelines. Two primary versions of SAM currently have been proposed, which we refer to as SAM and SAM2 for clarity.

The SAM comprises three core components: (1) an image encoder, (2) a prompt encoder, and (3) a mask decoder. The image encoder is based on the Vision Transformer (ViT) architecture, pretrained with Masked Autoencoders (MAE) [51] and is responsible for extracting high-level image features. The prompt encoder processes user-provided prompts—such as points, bounding boxes, or masks—into prompt embeddings. These embeddings are then fused with image features in the mask decoder to generate the final segmentation output. Given an image and a corresponding prompt, SAM processes them using the image and prompt encoders to obtain D-dimensional embeddings $Z^I$ and $Z^P$, respectively. These embeddings are then passed to the mask decoder to generate the predicted segmentation mask. The expected segmentation output in SAM can be represented as follows:

$$(m, c) = Decoder(Z^I | Z^P), \tag{1}$$

where $Z^P$ denotes the prompt features, $Z^I$ represents the features extracted by the image encoder, $m$ is the predicted binary segmentation mask, and $c$ is the associated confidence score.

SAM2 (illustrated in Figure 3) builds upon the architecture of SAM by introducing a streaming memory mechanism designed for real-time video and image processing. This enhancement is achieved through three additional modules: (1) memory attention, which conditions current frame features on past frame representations, predictions, and new prompts; (2) a memory encoder, which transforms the outputs of the image encoder and mask predictions into a latent memory representation for future use; and (3) a memory bank, which stores accumulated information about past predictions of the target object. These additions allow SAM2 to maintain contextual continuity and improve segmentation accuracy in sequential frames.

*3.2 Prompt Augmentation*

As shown in prior work, the quality and structure of prompts play a critical role in the segmentation performance of SAM2 [52]. In this study, we introduce a novel prompt augmentation strategy tailored for ART, wherein prior annotations (i.e., historical masks) are incorporated as prompts in combination with current bounding boxes. This approach aims to enrich prompt quality and robustness, thereby enhancing segmentation accuracy. Specifically, we use the mask from a prior MRI scan as an auxiliary prompt, integrated with a bounding box based on the current MRI slice. This design leverages anatomical consistency between treatment sessions to guide more precise segmentation. The overall pipeline of the proposed prompt augmentation strategy is illustrated in Figure 4 and consists of three key steps used in the training stage: (1) Prompt generation; (2) Prompt augmentation; and (3) Prompt selection, before feeding the augmented prompts into the prompt encoder of the proposed SAM2-Aug model.

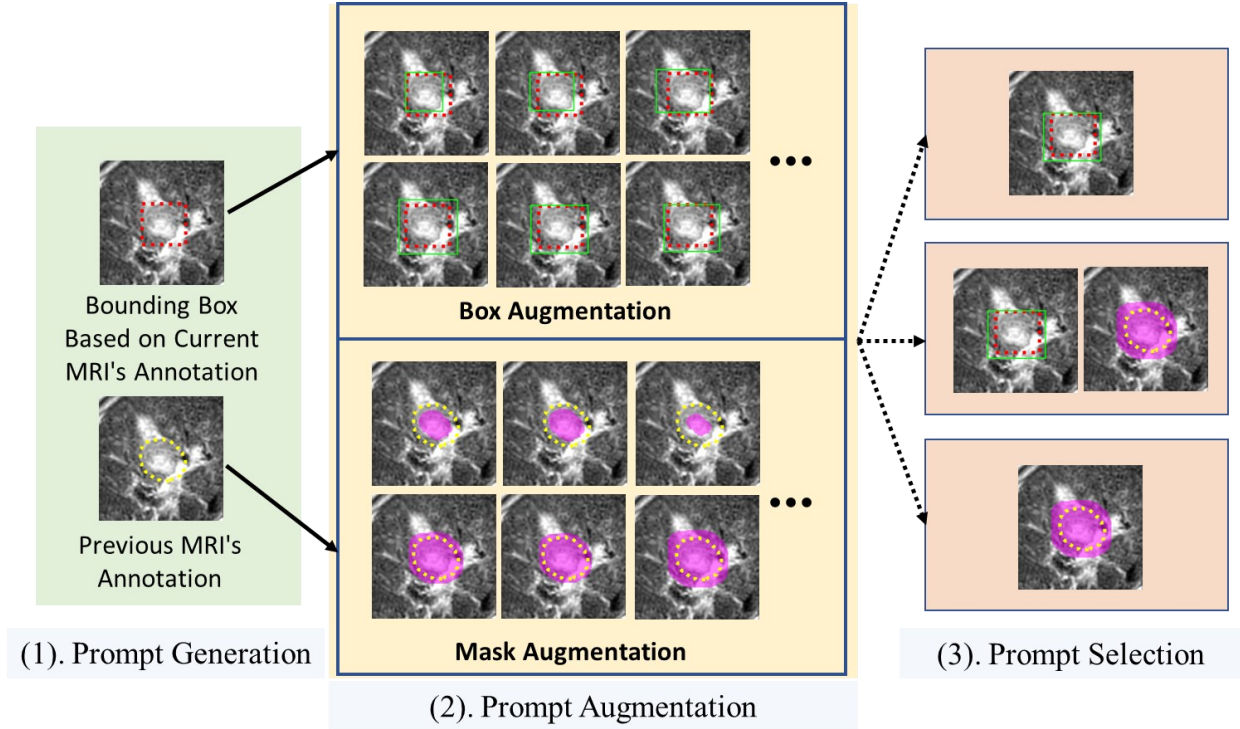

**Figure 4**. Pipeline of the proposed prompt augmentation strategy for adaptive radiation therapy (ART). The process consists of three main steps applied during network training: (1) Prompt generation, (2) Prompt augmentation, and (3) Prompt selection. Two types of augmentations—random bounding box expansion/contraction and random prior mask erosion/dilation—are employed to enhance robustness and leverage temporal consistency for improved tumor segmentation accuracy. For visualization, the prior annotated tumor boundary is shown in yellow, the bounding box from the current mask is in red, the augmented bounding box is in green, and the augmented prior mask is in purple. Dashed lines indicate potential selected scenarios following the stochastic prompt selection. For simplicity, the scenario where no prompts are selected for augmentation is omitted from the illustration.

In the training stage, first, the current MRI scan is retrieved and spatially aligned with the prior scan (with prior tumor mask) using rigid registration. These two prompts—the prior mask and the current bounding box—are initially combined to form the input for augmentation. The initial prompt set can be defined as follows:

$$Prompt_{init} = \{B_{cur}, M_{pre}\}, \qquad (2)$$

where the $B_{cur}$ is the current bounding box, and the $M_{pri}$ denotes the mask from the prior annotation.

Second, augmentation is applied to the prompts. For the mask prompt, morphological operations such as erosion and dilation are used to introduce spatial variations, allowing the model to learn from imperfect or noisy inputs while preserving the overall tumor shape. For the bounding box prompt, random expansions or contractions of 0-5 pixels are applied in all directions to simulate variability in user-defined inputs. It can be represented as follows:

$$Prompt_{Aug} = \{\mathcal{A}_b(B_{cur}), \mathcal{A}_m(M_{pri})\}, \tag{3}$$

where $\mathcal{A}_b$ denotes the random expansion or contraction on the initial bounding box, and $\mathcal{A}_m$ means the morphological operation (e.g., erosion or dilation) applied to the prior mask.

Third, a random selection mechanism is applied to the augmented mask and bounding box prompts. This stochastic strategy enhances the model's robustness to incomplete annotations, reflecting practical clinical scenarios where certain prompts may be unavailable. This design encourages improved generalization across diverse clinical inputs and varying annotation completeness. In summary, this proposed prompt augmentation strategy not only incorporates temporal information from prior annotations but also enhances the model's resilience to prompt variability, making it more robust for real-world ART applications.

During the testing stage, all bounding boxes used were similarly expanded or contracted by 0 to 5 pixels in each direction, for a randomly selected subset of the four directions (left, right, top, bottom). This augmentation simulates realistic variations in bounding box placement, mimicking potential variations encountered in clinical workflows. The prior tumor annotation was not augmented, and no dropout was performed, assuming physician-approved prior information will always be available in the ART workflow.

*3.3 Input data augmentation and fine-tuning SAM2*

In ART, both current and prior MRI scans are available. To effectively leverage this temporal (previous) information, we construct a three-channel input to SAM2 by concatenating the current MRI, the prior MRI, and the corresponding prior tumor annotation. This design introduces anatomical and contextual priors into the model. For instance, for adaptive fraction *n* of MR-LINAC-based ART, we are using the MRI scan and tumor annotation of fraction (*n-1*) as prior. For the first fraction, the MRI scan and tumor annotation from the initial treatment simulation and planning were used. The use of the *immediate prior* allows the incorporation of the most relevant prior knowledge, especially in cases where the tumor morphology may undergo rapid changes.

Figure 5 presents examples from three different patients, illustrating the integration of the current MRI, the prior MRI, and the associated prior tumor annotation on the same slice. The 'ground-truth' tumor contours are overlaid in yellow on each MR image. This input data augmentation approach enables SAM2 to utilize both historical and current information, improving its ability to localize and delineate tumors in ART with higher accuracy and consistency.

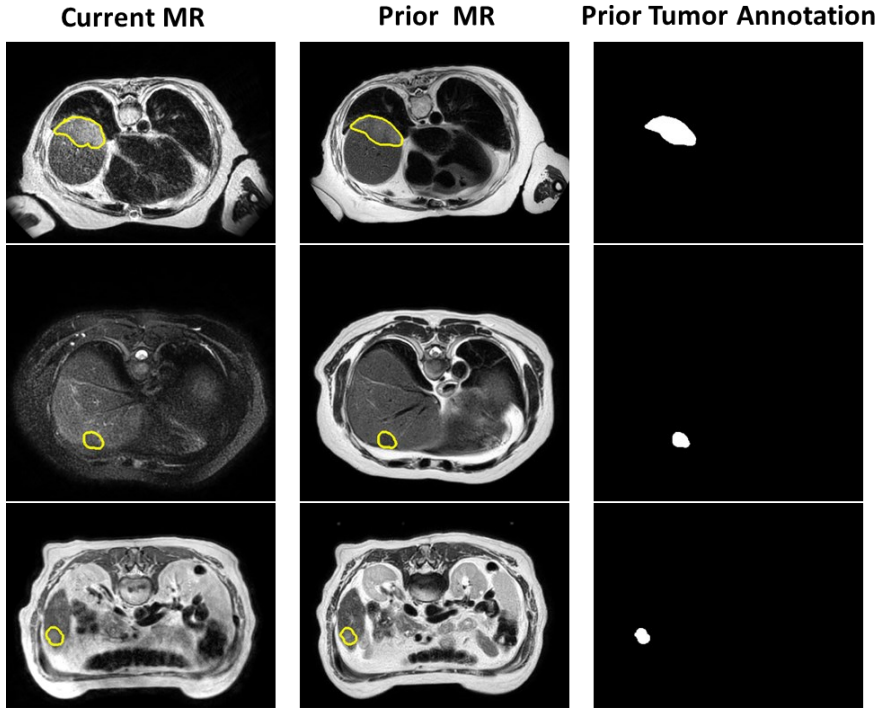

**Figure 5**. The input data augmentation strategy. The current MRI, prior MRI, and corresponding prior tumor annotation are concatenated to form a three-channel input to SAM2, integrating anatomical and contextual priors to enhance tumor segmentation in adaptive radiation therapy. For visualization, tumor boundaries are overlaid on the MRI slices. The examples demonstrate a strong spatial consistency between current and prior tumors within the same case.

As the original SAM2 model was trained on natural images and videos, a significant domain gap exists when applying it to medical image segmentation tasks. Therefore, fine-tuning SAM2 is essential to enhance its performance in the medical domain [29]. In this study, we adapt SAM2 for tumor segmentation in ART. Both the image encoder, prompt encoder, and mask decoder are fine-tuned jointly via backpropagation. The resulting model—referred to as SAM2-Aug—enhances tumor segmentation in ART by leveraging temporally enriched inputs and robust prompt augmentation within the SAM2 framework.

## 4. Dataset and Implementation

### 4.1 Datasets

To evaluate the effectiveness of SAM2-Aug, we curated three MRI datasets derived from ART workflows. The first dataset, *One-Seq-Liver*, consists of 115 MRI scans from 31 patients diagnosed with liver cancer, all acquired using the same T2-weighted MultiVane eXtended Dynamic sequence (T2-MVXD). There are 31 additional simulation scans from the 31 cases, which are the very first image/annotation of each patient case and will serve as the prior information for the first adaptive fraction (same for the other datasets). All patients were treated on a 1.5T MR-LINAC (Unity, Elekta AB, Stockholm, Sweden), with 1~7 fractions per patient. Such fractions only include those where contours are modified for adaptive-to-shape treatments [53]. In total, 167 internal target volumes (ITVs) were manually delineated by experienced radiation oncologists involved in daily treatments.

The second dataset, *Mix-Seq-Abdomen*, comprises 88 fractional MRI scans from 28 MR-LINAC patients with abdominal cancers, including liver, pancreatic, and renal tumors. This dataset incorporates multiple MRI sequences—T2-MVXD, T2-MVXD with fat suppression (T2-MVXD-SPIR), T1-weighted 3D Vane XD (T1-3D-VaneXD), and T2-weighted fat-suppressed turbo spin echo imaging (T2-FS)—used across different treatment fractions (2~6 fractions per patient besides simulation). A total of 156 ITVs were annotated by radiation oncologists for this cohort.

The third dataset, *Mix-Seq-Brain*, consists of 86 MRI scans with 1~5 fractions from 37 MR-LINAC patients with brain tumors. This dataset also includes multiple T1- or T2-weighted MRI sequences (T2-FLAIR, T1, and T1 with contrast). A total of 97 clinical target volumes (CTVs) were annotated.

A detailed summary of the three datasets is presented in Table 1. Notably, only the One-Seq-Liver dataset was used for training, while the Mix-Seq-Abdomen and Mix-Seq-Brain datasets were exclusively reserved for testing, to assess the robustness and generalizability of SAM-Aug2.

**Table 1.** Clinical characteristics of patient cases in the three datasets. Simulation scans are excluded from the counts since they can only be used as prior information.

| Dataset Name | Target Type | No. of Patient Cases | No. of Scans | No. of Targets contoured | Mean Target Vol (mm$^3$) | MR Sequences |
|---|---|---|---|---|---|---|
| One-Seq-Liver | ITV | 31 | 115 | 167 | 45,021 | T2-MVXD |
| Mix-Seq-Abdomen | ITV | 28 | 88 | 156 | 26,664 | T2-MVXD, T2-MVXD-SPIR, T1-3D-VaneXD, and T2-FS |
| Mix-Seq-Brain | CTV | 37 | 86 | 97 | 156,772 | T2-FLAIR, T1, T1 Contrast |

The One-Seq-Liver dataset was divided into training, validation, and testing subsets comprising 19, 5, and 7 patients, corresponding to 69, 17, and 29 MRI scans, respectively, as detailed in Table 2.

**Table 2.** The data splitting information for the One-Seq-Liver dataset.

| One-Seq-Liver | No. of Training Scans | No. of Validation Scans | No. of Testing Scans | No. of Targets in Testing Set | Mean Target Vol in Testing Set (mm$^3$) |
|---|---|---|---|---|---|
| No. of Cases | 19 | 5 | 7 | 52 | 15465 |
| No. of Scans | 69 | 17 | 29 | | |

*4.2 Implementation details*

All experiments involving SAM2-Aug were conducted using the PyTorch framework on an Ubuntu system equipped with a single NVIDIA RTX 4090 GPU. The pretrained SAM2-Tiny model served as the foundation for fine-tuning. During training, we employed a batch size of 4 and used the AdamW optimizer with an initial learning rate of $6 \times 10^{-6}$ and a weight decay of 0.01. The model was fine-tuned for 5 epochs, as additional epochs did not yield further improvements on the validation set. A combination of Dice Similarity Coefficient (Dice) loss and cross-entropy loss was employed as the training objective. The final model was selected based on the highest Dice score achieved on the validation set.

*4.3 Comparison methods*

Registration-based, convolution-based (UNet [12], nn-UNet [18], SegResNet [20], and MedNeXt [17]), transformer-based (Swin-UNETR [15] and SegFormer [19]) and SAM-based (MedSAM [22], SAM-Med2D [29], SAM-Med3D [30] and SAM-Med3D-Turbo [30]) were compared with SAM2-Aug as baseline models.

For the registration-only approach, we employed the ANTs package[1] to perform rigid registration. The first frame of each video sequence was treated as the moving image and registered to all subsequent frames. The annotated mask from the first frame was then propagated across the sequence, serving as pseudo-annotations for the remaining frames.

As for the CNN and Transformer-based methods, training for these models was conducted in 3D using input patch volumes of size 128 × 128 × 128 × 3 (for three-channel inputs, similar to SAM2-Aug), sampled via a random cropping strategy, with a batch size of 2. Data augmentation techniques included random Gaussian noise (applied with a probability of 0.1 and a standard deviation of 0.01) and intensity scaling within the range of 0.9 to 1.1. Each model received the current MR image, the prior MR image, and its associated prior annotation as input, as relying solely on the current MR image resulted in suboptimal performance. The AdamW optimizer was employed with an initial learning rate of 0.001 and a weight decay of $1 \times 10^{-5}$. A cosine annealing schedule was applied to gradually reduce the learning rate to a minimum of $1 \times 10^{-7}$. All models were trained for 100 epochs on a single NVIDIA RTX 4090 GPU. The final model was selected according to the highest Dice score achieved on the validation dataset of One-Seq-Liver. While task-specific CNN and transformer models require no prompts, SAM-based methods depend on interactive inputs. For fairness, we followed the original settings by using bounding boxes for MedSAM and SAM-Med2D, ten points for SAM-Med3D and SAM-Med3D-Turbo, and both bounding boxes and masks for SAM2 and SAM2-Aug.

*4.4 Evaluation Metrics*

Segmentation quality is assessed using four widely adopted metrics: Dice, Normalized Surface Dice (NSD), 95th Percentile Hausdorff Distance (HD95), and Average Surface Distance (ASD). Together, these metrics offer a comprehensive evaluation of volumetric overlap, surface conformity, and boundary accuracy.

---

[1] https://github.com/ANTsX/ANTs

Specifically, the Dice measures the overlap between the predicted segmentation volume and the 'ground truth', defined as:

$$Dice(G, P) = 2 \frac{|G \cap P|}{|G| \cup |P|}, \quad (4)$$

where the G and P indicate the 'ground-truth' and predicted masks, respectively.

NSD evaluates the surface-level agreement between predicted and 'ground-truth' segmentations. It calculates the proportion of surface voxels from each that lie within a specified tolerance (τ) of the other:

$$NSD = \frac{|\{x \in S_G : d(x, S_P) \leq \tau\}| + |\{x \in S_P : d(x, S_G) \leq \tau\}|}{|S_G| + |S_P|}, \quad (5)$$

where $S_G$ and $S_P$ are surfaces of the 'ground-truth' and predicted segmentations, respectively; and |·| measures the counts of voxels. $d(x, S)$ is the minimum distance from the point $x$ to surface $S$. The tolerance threshold $\tau$ was set to 0.5 in this study.

HD95 is used to evaluate the boundary accuracy of a segmentation by measuring the distance between the surfaces of the predicted and 'ground- truth' segmentations. HD95 is defined as:

$$HD_{95}(S_G, S_P) = max\left(P_{a \in S_G} min_{b \in S_P} d(a,b), P_{b \in S_P} min_{a \in S_G} d(b,a)\right), \quad (6)$$

where $P$ means 95$^{th}$ percentile, and $d(a, b)$ represents the Euclidean distance between boundary points $a$ and $b$.

ASD measures the average of the shortest distances between the surface points of the predicted segmentation and the 'ground truth', providing an overall sense of how closely the two surfaces align. The definition of ASD can be represented as:

$$ASD(S_G, S_P) = \frac{1}{|S_G| + |S_P|} \left( \sum_{x \in S_G} d(x, S_P) + \sum_{x \in S_P} d(x, S_G) \right), \quad (7)$$

where $d(x, S)$ is the minimum Euclidean distance from point $x$ to surface $S$.

## 5. Experimental results

*5.1 Comparison between SAM2-Aug and the other segmentation methods*

(1) Performance on the One-Seq-Liver dataset

We evaluated SAM2-Aug on the One-Seq-Liver test set and compared it with state-of-the-art segmentation methods (Table 3). Among all methods on One-Seq-Liver dataset, SAM2-Aug achieved the best overall performance with a mean Dice score of 0.86, mean NSD of 0.58, mean HD95 of 1.98, and mean ASD of 0.60, indicating better accuracy in both volumetric overlap and boundary alignment. In comparison, the best-performing interactive-based method, MedSAM, achieved a Dice score of 0.83, while conventional CNN-based models such as SegResNet and nn-UNet reported Dice scores of 0.82 and 0.80, respectively. Transformer-based models such as SwinUNETR and SegFormer underperformed slightly, especially in boundary-related metrics like HD95 and ASD. In addition, the baseline SAM2 performed worse than SAM2-Aug, highlighting the importance of our prompt augmentation strategy. Notably, the registration-only method outperformed both convolutional and transformer-based approaches. We hypothesize that deep learning models may be more susceptible to overfitting on the training data, and that even minor variations in tumor segmentation predictions can substantially impact evaluation metrics, especially for small or irregularly shaped lesions.

**Table 3.** Comparison of tumor segmentation performance on the test set of the One-Seq-Liver dataset. The arrows are pointing in the direction of improved accuracy. The best values are shown in bold. The best values are shown in bold.

| Method | Type | DICE↑ | NSD↑ | HD95↓[vol] | ASD↓[vol] |
|---|---|---|---|---|---|
| Registration-only | Rigid | 0.83±0.13 | 0.53±0.25 | 2.13±1.68 | 0.64±0.52 |
| UNet | CNN | 0.70±0.19 | 0.29±0.18 | 221.18±102.12 | 70.08±56.06 |
| nn-UNet | CNN | 0.80±0.12 | 0.41±0.16 | 7.39±26.98 | 4.32±15.96 |
| SegResNet | CNN | 0.82±0.11 | 0.44±0.18 | 2.41±1.58 | 0.70±0.50 |
| MedNeXt | CNN | 0.81±0.12 | 0.44±0.19 | 18.30±57.94 | 2.45±5.88 |
| SwinUNETR | Transformer | 0.79±0.14 | 0.44±0.22 | 169.43±164.43 | 35.70±35.45 |
| SegFormer | Transformer | 0.76±0.13 | 0.32±0.11 | 52.83±86.82 | 8.57±12.65 |
| MedSAM | SAM-based | 0.83±0.09 | 0.53±0.12 | 2.35±1.01 | 0.63±0.26 |
| SAM-Med2D | SAM-based | 0.71±0.20 | 0.40±0.15 | 5.34±3.28 | 1.68±1.05 |
| SAM-Med3D | SAM-based | 0.12±0.10 | 0.02±0.02 | 52.33±20.66 | 23.40±10.57 |
| SAM-Med3D-Turbo | SAM-based | 0.18±0.12 | 0.04±0.03 | 27.73±10.56 | 11.61±5.61 |
| SAM2 (without prompt augmentation) | SAM-based | 0.78±0.13 | 0.49±0.18 | 22.37±41.96 | 5.06±8.96 |
| SAM2-Aug | SAM-based | **0.86±0.08** | **0.58±0.12** | **1.98±0.66** | **0.60±0.23** |

Figure 6 presents qualitative comparisons across multiple cases between the proposed SAM2-Aug method and existing approaches, including MedSAM, SAM-Med2D, SegResNet, and Swin-UNETR. 'Ground-truth' annotations are delineated by red contours. The regions highlighted by yellow bounding boxes are zoomed in to provide a detailed view, where predicted tumor segmentations are shown as green overlays on the cropped MR images. Notably, SAM2-Aug demonstrates superior segmentation performance, particularly in preserving tumor shape and boundary accuracy, compared to the other methods.

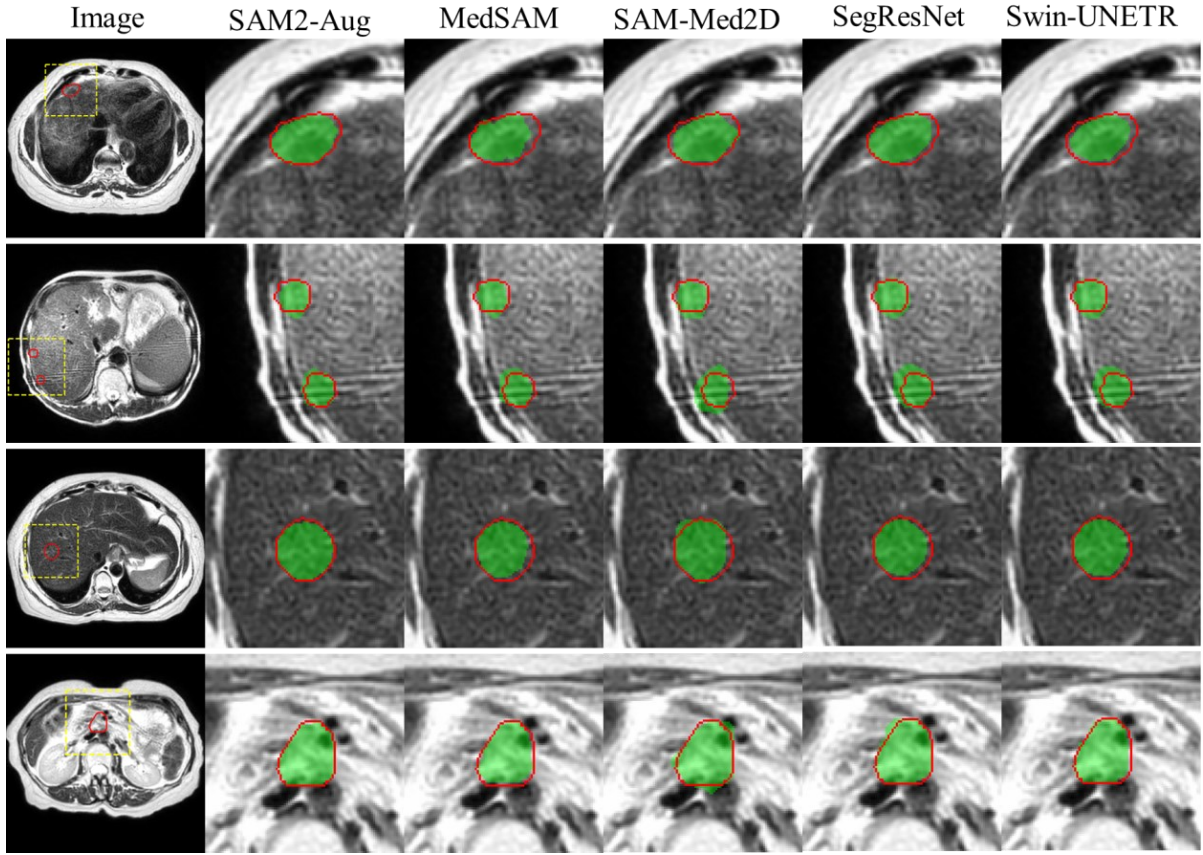

**Figure 6**. Qualitative comparison of tumor segmentation. SAM2-Aug is compared with MedSAM, SAM-Med2D, SegResNet, and Swin-UNETR. 'Ground-truth' boundaries are shown in red, with green overlays indicating predictions. Yellow boxes mark zoomed-in regions, highlighting SAM2-Aug's improved tumor delineation accuracy.

(2) Generalizability testing on Mix-Seq-Abdomen and Mix-Seq-Brain

To evaluate the generalizability of SAM2-Aug across varied MRI protocols and tumor types, we tested it on the Mix-Seq-Abdomen dataset, focusing on ITV segmentation under high inter-sequence variability. As shown in Table 4, SAM2-Aug achieved the best performance among SAM-based methods, with a mean Dice of 0.89, mean NSD of 0.53, mean HD95 of 2.14, and mean ASD of 0.67. In contrast, methods like MedSAM, SAM-Med2D, and SAM2 reached Dice scores of 0.84 but underperformed in boundary accuracy. 3D models such as SAM-Med3D and SAM-Med3D-Turbo showed poor generalizability with Dice scores below 0.15.

**Table 4**. Comparison of tumor segmentation performance on the Mix-Seq-Abdomen dataset. The arrows are pointing in the direction of improved accuracy. The best values are shown in bold.

| Method | Type | DICE↑ | NSD↑ | HD95↓[vol] | ASD↓[vol] |
|---|---|---|---|---|---|
| Registration-only | Rigid | 0.83±0.14 | 0.37±0.24 | 2.67±2.22 | 0.95±0.80 |
| UNet | CNN | 0.81±0.13 | 0.30±0.14 | 74.47±93.15 | 13.49±16.25 |
| nn-UNet | CNN | 0.82±0.14 | 0.31±0.17 | 11.51±34.65 | 2.41±4.48 |
| SegResNet | CNN | 0.83±0.14 | 0.33±0.19 | 4.67±16.29 | 2.06±9.632 |
| MedNeXt | CNN | 0.81±0.13 | 0.31±0.18 | 11.49±28.99 | 2.61±2.661 |
| SwinUNETR | Transformer | 0.58±0.22 | 0.16±0.11 | 296.33±58.01 | 147.68±63.78 |
| SegFormer | Transformer | 0.72±0.17 | 0.20±0.10 | 144.91±89.88 | 39.27±36.69 |
| MedSAM | SAM-based | 0.84±0.08 | 0.42±0.14 | 2.96±1.52 | 0.84±0.39 |
| SAM-Med2D | SAM-based | 0.84±0.09 | 0.40±0.11 | 3.66±2.03 | 1.15±0.66 |
| SAM-Med3D | SAM-based | 0.06±0.12 | 0.01±0.02 | 105.21±36.80 | 59.66±33.70 |
| SAM-Med3D-Turbo | SAM-based | 0.14±0.15 | 0.02±0.02 | 105.24±59.46 | 47.03±35.16 |
| SAM2 (without prompt augmentation) | SAM-based | 0.84±0.11 | 0.45±0.18 | 15.81±29.05 | 3.33±5.70 |
| SAM2-Aug | SAM-based | **0.89±0.06** | **0.53±0.14** | **2.14±1.41** | **0.67±0.39** |

Figure 7 illustrates tumor segmentation results on the Mix-Seq-Abdomen dataset using various methods. Visually, SAM2-Aug consistently produces segmentations that closely align with the red 'ground-truth' boundaries, showing precise tumor localization and accurate delineation of irregular shapes. In contrast, MedSAM and SAM-Med2D often suffer from boundary mismatches or slight over-segmentation, especially in low-contrast or small tumor regions. The CNN-based SegResNet and transformer-based SegFormer exhibit limited adaptability in complex anatomical contexts, frequently under-segmenting or missing subtle tumor extensions. Overall, SAM2-Aug demonstrates superior visual performance with smoother, anatomically consistent boundaries, highlighting its robustness across varying tumor sizes, shapes, and imaging conditions.

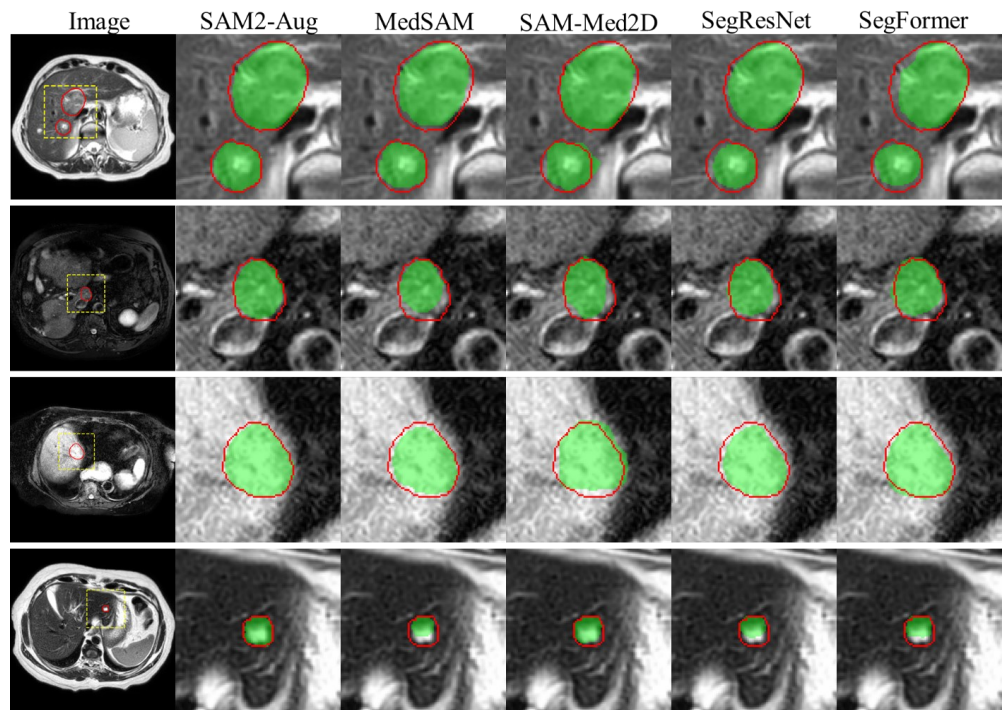

**Figure 7**. Qualitative comparison of tumor segmentation in the ITV region on the Mix-Seq-Abdomen dataset. Each row represents a different case. The first column shows the full MR slice with a yellow bounding box indicating the zoomed-in region. Subsequent columns display zoomed-in predictions from different methods: SAM2-Aug, MedSAM, SAM-Med2D, SegResNet,

and SegFormer (left to right). 'Ground-truth' tumor boundaries are shown in red, and predicted segmentations are overlaid in green.

In addition, we also tested it on the Mix-Seq-Brain dataset, which contains brain tumors with distinct imaging characteristics compared to the abdominal datasets. As shown in Table 5, SAM2-Aug achieved the highest performance among all compared methods, with a mean Dice score of 0.90, mean NSD of 0.44, mean HD95 of 3.96, and mean ASD of 1.15. Compared to the original SAM2, SAM2-Aug improved the Dice score by 0.02 and significantly reduced boundary-related errors. Additionally, SAM2-Aug outperformed the rigid registration baseline by 0.05 in Dice and showed marked improvements in all surface-based metrics. These results underscore the strong generalization ability of SAM2-Aug across anatomical regions and tumor types, highlighting its potential for broader clinical applicability.

**Table 5.** Comparison of CTV segmentation performance on the brain tumor dataset. Arrows are pointing in the direction of improved accuracy. The best values are shown in bold.

| Method | Type | DICE↑ | NSD↑ | HD95↓[vol] | ASD↓[vol] |
|---|---|---|---|---|---|
| Registration-only | Rigid | 0.85±0.16 | 0.36±0.24 | 4.98±5.17 | 1.73±3.25 |
| UNet | CNN | 0.75±0.20 | 0.20±0.15 | 137.73±117.97 | 27.22±27.83 |
| nn-UNet | CNN | 0.47±0.26 | 0.06±0.07 | 281.83±71.62 | 148.75±71.13 |
| SegResNet | CNN | 0.80±0.20 | 0.26±0.18 | 25.49±24.95 | 8.10±10.75 |
| MedNeXt | CNN | 0.63±0.28 | 0.17±0.14 | 11.08±6.85 | 3.55±3.35 |
| SwinUNETR | Transformer | 0.73±0.25 | 0.22±0.16 | 13.84±27.76 | 5.28±10.31 |
| SegFormer | Transformer | 0.77±0.20 | 0.17±0.10 | 9.15±12.63 | 2.60±3.19 |
| MedSAM | SAM-based | 0.76±0.18 | 0.26±0.10 | 7.84±3.99 | 2.19±1.22 |
| SAM-Med2D | SAM-based | 0.84±0.09 | 0.31±0.11 | 8.32±4.57 | 2.38±1.24 |
| SAM-Med3D | SAM-based | 0.16±0.17 | 0.02±0.02 | 53.30±25.49 | 24.22±15.48 |
| SAM-Med3D-Turbo | SAM-based | 0.34±0.19 | 0.04±0.03 | 53.88±35.07 | 19.29±12.71 |
| SAM2 (without prompt augmentation) | SAM-based | 0.88±0.08 | 0.39±0.17 | 9.86±16.21 | 2.34±2.87 |
| SAM2-Aug | SAM-based | **0.90±0.07** | **0.44±0.15** | **3.96±2.80** | **1.15±0.84** |

*5.2 Ablation studies*

(1) SAM2-Aug with different input data augmentation

Table 6 presents the ITV segmentation performance of SAM2-Aug under different input data configurations on the first One-Seq-Liver dataset. Using only the current MR (curMR) yielded the lowest performance. Incorporating the prior MR (priMR) led to a 0.05 (in absolute terms, same in the following) improvement in Dice, while adding the prior segmentation (priSeg) increased Dice performance by 0.07, indicating that prior images/annotations offer valuable context for current tumor segmentation. The highest performance was achieved when curMR, priMR, and priSeg were combined, resulting in a 0.11 improvement in Dice and a 0.14 increase in NSD, along with substantial reductions in HD95 and ASD compared to using curMR alone.

Table 6. The comparison between SAM2-Aug results using different inputs for data augmentation on the One-Seq-Liver Dataset. Here, the curMR, priMR, and priSeg denote the current MR, the prior MR, and the tumor annotation of the prior MR. Arrows are pointing in the direction of improved accuracy. The best values are shown in bold.

| Input Data Type | DICE↑ | NSD↑ | HD95↓[vol] | ASD↓[vol] |
|---|---|---|---|---|
| curMR | 0.75±0.12 | 0.44±0.08 | 4.82±1.16 | 1.46±0.41 |
| curMR+priMR | 0.80±0.12 | 0.49±0.13 | 2.84±0.86 | 0.94±0.41 |
| curMR+priSeg | 0.82±0.10 | 0.52±0.11 | 2.67±0.72 | 0.85±0.31 |
| curMR+priMR+priSeg | **0.86±0.08** | **0.58±0.12** | **1.98±0.66** | **0.60±0.23** |

(2) SAM2-Aug with different prompt augmentation

Table 7 summarizes the performance of SAM2-Aug on ITV segmentation using different prompt types—mask, bounding box (BBox), and their combinations—with and without prompt augmentation. Four main observations can be made: (a) BBox prompts outperformed mask prompts. (b) Augmenting the BBox prompt further improved Dice by 0.02, demonstrating the value of prompt diversity. (c) Combining BBox and mask without augmentation provided a modest improvement over mask alone, but underperformed compared to BBox alone. (d) The best results were achieved when both BBox and mask prompts were augmented, yielding a 0.1 Dice improvement over the baseline with the most favorable boundary accuracy. These findings underscore the critical role of prompt augmentation—particularly with BBox—in enhancing segmentation accuracy.

Table 7. Comparison of ITV segmentation performance using different types of prompts—mask, bounding box, and their combinations—with and without prompt augmentation on the One-Seq-Liver Dataset. Arrows are pointing in the direction of improved accuracy. The best values are shown in bold.

| Prompt Type | DICE↑ | NSD↑ | HD95↓[vol] | ASD↓[vol] |
|---|---|---|---|---|
| Mask | 0.73±0.25 | 0.41±0.22 | 29.08±78.27 | 14.38±40.06 |
| BBox | 0.82±0.10 | 0.52±0.13 | 2.78±1.04 | 0.86±0.39 |
| AugBBox | 0.84±0.08 | 0.55±0.12 | 2.25±0.74 | 0.67±0.25 |
| BBox+Mask | 0.76±0.12 | 0.43±0.09 | 3.83±0.83 | 1.33±0.48 |
| AugBBox+AugMask (SAM2-Aug) | **0.86±0.08** | **0.58±0.12** | **1.98±0.66** | **0.60±0.23** |

To further evaluate the robustness of different prompt augmentation strategies under variations in bounding box (BBox) specification, we randomly expanded the BBox by 1 to 10 pixels in one to four directions on the first and second datasets during testing. As shown in Figure 8, the proposed prompt augmentation strategy yielded more stable segmentation performance across both datasets, demonstrating enhanced tolerance to BBox variability. Furthermore, the use of mask prompts proved especially beneficial for segmenting small tumors. This is evidenced by the improved performance on the first dataset, where the mean ITV volume (15,465 mm³) is considerably smaller than that of the second dataset (26,664 mm³).

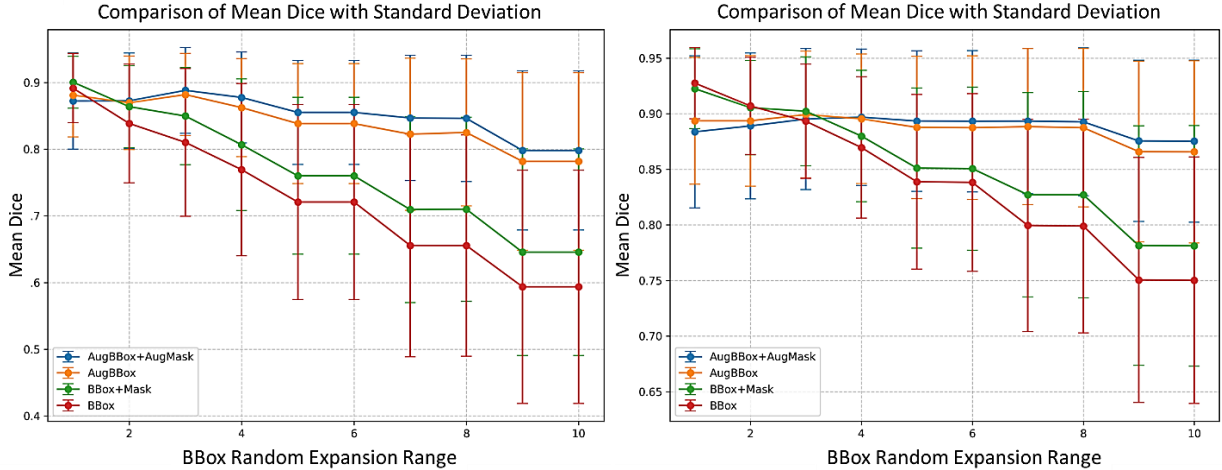

**Figure 8**. Comparison of ITV segmentation in terms of mean Dice scores, with and without prompt augmentation on the first and second datasets. This figure illustrates the impact of bounding box random expansion—applied in all directions randomly within a range of 1 to 10 pixels—on segmentation performance across the One-Seq-Liver (left) and Mix-Seq-Abdomen (right) datasets during testing. The results highlight the effectiveness of prompt augmentation in mitigating variability in user interaction when providing the bounding box.

(3) SAM2-Aug with different fine-tuned modules

To assess the contribution of different components within SAM2-Aug, we conducted an ablation study by fine-tuning individual and combined modules, including the mask decoder (MskDec), image encoder (ImgEnc), prompt encoder (PrmpEnc), and all modules together. As shown in Table 7, fine-tuning only the MskDec or ImgEnc led to moderate improvements, with average Dice scores of 0.81 and 0.83 on the first dataset, respectively. Joint fine-tuning of ImgEnc and MskDec further improved performance (average Dice: 0.84). The best segmentation results were achieved by fine-tuning ImgEnc, MskDec, and PrmpEnc jointly, yielding the highest Dice (0.86) and lowest HD95 and ASD. Interestingly, fine-tuning all modules did not outperform this configuration, suggesting that selectively updating key modules—rather than the full model—may be more effective for adapting SAM2-Aug to medical segmentation tasks.

**Table 7**. Comparison of segmentation performance between different fine-tuning strategies of SAM2-Aug on the One-Seq-Liver Dataset. The arrows are pointing in the direction of improved accuracy. The best values are shown in bold.

| Fine-tuning Module | Parameters (M) | DICE↑ | NSD↑ | HD95↓[vol] | ASD↓[vol] |
|---|---|---|---|---|---|
| MskDec | 4.2 | 0.81±0.12 | 0.53±0.14 | 2.71±1.05 | 0.87±0.44 |
| ImgEnc | 27.2 | 0.83±0.10 | 0.54±0.11 | 2.53±0.82 | 0.78±0.32 |
| ImgEnc + MskDec | 31.4 | 0.84±0.09 | 0.55±0.11 | 2.43±0.75 | 0.74±0.28 |
| ImgEnc + MskDec + PrmpEnc | 31.4 | **0.86±0.08** | **0.58±0.12** | **1.98±0.66** | **0.60±0.23** |
| All modules | 38.9 | 0.85±0.09 | 0.57±0.11 | 2.19±0.68 | 0.66±0.26 |

## 6. Discussion

In this study, we aimed to explore how to adapt SAM2 for ART by leveraging prior knowledge from previous treatments. We found that simple input and prompt augmentations are effective strategies for improving tumor segmentation performance in ART. However, there are still several important limitations to consider in the future:

(1) Use of one/few prompts during inference for tumor segmentation

While the streaming memory mechanism in SAM2 enables the model to leverage temporal relationships between frames—allowing segmentation with only a few prompts, or even a single prompt for one or a few slices by treating each MR volume as a video—this approach proved less effective for tumor segmentation in our ART setting. Specifically, we observed that using only partial prompts during inference did not yield satisfactory segmentation performance. This limitation may stem from the insufficient contextual support provided by the memory module, which fails to offer adequate guidance for accurate segmentation in the current frame. Unlike natural video data, where strong and continuous temporal coherence exists, the spatial and temporal correlations in 3D medical imaging for ART are less consistent and may not be as informative. Therefore, relying solely on temporal memory across slices may not be ideal in this context.

A potential approach involves dynamically generating prompts for subsequent slices by integrating the current segmentation prediction with annotations from adjacent slices. This strategy leverages both present and historical contextual information to iteratively refine prompts, thereby emulating a one- or few-shot prompting process during inference. In addition, tumor-related features extracted from previous MRI slices can be stored in a memory bank to further enhance model performance. These memory representations can be leveraged to inform and guide feature updates in the current MRI slice, thereby enabling more robust tumor tracking and segmentation across consecutive slices. Within this framework, the need for explicit bounding box prompts in each new MRI scan may be eliminated. The sequential MRI volumes of a patient can be interpreted as a temporal sequence, analogous to a video, where each time step corresponds to a 3D volume.

(2) SAM 2 with 3D adapters

In this study, we fine-tuned the original SAM2 model using limited data in the ART setting. Although the results demonstrate the effectiveness of slice-by-slice training and inference, this approach does not account for the 3D contextual information inherent in tumor structures. Recent studies [48, 54, 55] have highlighted the benefits of incorporating 3D adaptation techniques into SAM-based architectures, which allow the integration of domain-specific medical knowledge for more accurate 3D medical image segmentation.

Introducing a 3D adapter into SAM2 could be a promising direction for enhancing performance in volumetric segmentation tasks. However, determining where and how to insert such an adapter within the SAM2 framework remains an open research question. Moreover, a critical challenge lies in balancing segmentation accuracy with computational efficiency, particularly for time-sensitive applications such as real-time adaptive radiation therapy. Exploring optimal strategies for 3D integration while maintaining practical runtime constraints is an important area for future investigation.

(3) Exploring additional prompt augmentation strategies for SAM2

The nnInteractive framework [56] introduced 3D promptable segmentation using nn-UNet, employing a variety of prompt types—including points, scribbles, boxes, and lasso—by issuing 2D interactions to generate full 3D segmentations. Inspired by this approach, leveraging a small number of informative 2D prompts to generate 3D prompts could be a more effective strategy for tumor segmentation using SAM2 in ART.

Furthermore, combining multiple prompt types with tailored augmentation strategies—based on well-defined criteria or assumptions—may enhance the robustness and accuracy of 3D prompt generation. For instance, shape-invariant transformations could be applied for mask prompts, while information-theoretic measures such as maximum entropy [42] could guide point prompt augmentation.

In future work, we aim to explore advanced 3D prompt augmentation techniques in conjunction with adapter-based fine-tuning strategies. This direction holds promise for improving 3D tumor segmentation performance in ART through more expressive and context-aware prompting mechanisms.

## 7. Conclusion

In this study, we propose two novel and effective augmentation strategies for inputs and prompts in ART. Our results show that integrating the current MRI, prior MRI, and the corresponding annotation of the prior MRI enhances tumor segmentation by providing essential prior knowledge, such as tumor location and shape, leading to more accurate results. We also demonstrate that prompt augmentation for the bounding box and mask improves generalization, robustness, and overall performance in tumor segmentation. These strategies were implemented and evaluated on SAM2-Aug, with fine-tuning of the original SAM2 model. Experimental results confirm that the proposed methods enhance tumor segmentation performance, offering promising potential for advancing ART-based automatic tumor segmentation in clinical applications.

**Acknowledgement**

The study was supported by the US National Institutes of Health (R01 CA240808, R01 CA258987, R01 EB034691, and R01 CA280135).